\documentclass[preprint,preprintnumbers, prd, floatfix,  superscriptaddress,
nofootinbib] {revtex4}
\usepackage{graphicx}
\usepackage{epsfig}
\usepackage{bm}
\usepackage{amssymb}
\usepackage{float}
\usepackage{amsmath}
\usepackage{dcolumn}
\usepackage{cancel}
\usepackage[colorlinks]{hyperref}
\usepackage[usenames,dvipsnames]{color}
\usepackage{epstopdf}
\hypersetup{
     breaklinks=true,
    pdfstartview={FitH},    colorlinks=true,       % false: boxed links; true: colored links
    linkcolor=blue,          % color of internal links
    citecolor=red,        % color of links to bibliography
    filecolor=magenta,      % color of file links
    urlcolor=blue,           % color of external links
    anchorcolor=green,      % Color for anchor text
    linktocpage=true
}

\def\doi{http://doi.org}

\begin{document}

 \title{Observational constraints on running vacuum model}
 
\author{Jin-Jun Zhang}
%\email{zhangjinjun@sxnu.edu.cn}
\affiliation{School of Physics and Information Engineering, Shanxi Normal University, Linfen 041004}

\author{Chung-Chi Lee}
%\email{ccl51@damtp.cam.ac.uk}
\affiliation{DAMTP, Centre for Mathematical Sciences, University of Cambridge, Wilberforce Road, 
Cambridge CB3 0WA}

\author{Chao-Qiang Geng}
\email{geng@phys.nthu.edu.tw}
%\affiliation{Chongqing University of Posts \& Telecommunications, Chongqing, 400065,
%China}
\affiliation{Department of Physics, National Tsing Hua University, Hsinchu 300}
 \affiliation{National Center for Theoretical Sciences, Hsinchu 300}
\affiliation{School of Physics and Information Engineering, Shanxi Normal University, Linfen 041004}

\begin{abstract}
We investigate the power spectra of the CMB temperature and matter density in the running vacuum model (RVM) with the time-dependent cosmological constant of $\Lambda = 3 \nu H^2 + \Lambda_0$, where $H$ is the Hubble parameter. In this model, dark energy decreases in time and decays to both matter and radiation. By using the Markov chain Monte Carlo method, we constrain the model parameter $\nu$ as well as the cosmological observables. Explicitly, we obtain $\nu \leq 1.54 \times 10^{-4}$  (68\% confidence level) in the RVM with the best-fit $\chi^2_{\mathrm{RVM}} = 13968.8$, which is slightly smaller than $\chi^2_{\Lambda \mathrm{CDM}} = 13969.8$ in the $\Lambda$CDM model of 
$\nu=0$.
 \end{abstract}

\maketitle

\section{Introduction} \label{sec:introduction}

One of the most important recent cosmological observations is that our universe is undergoing a late-time accelerating expansion phase,
realized by introducing dark energy~\cite{Copeland:2006wr}.
Among the many possible dark energy scenarios, the simplest one is the $\Lambda$CDM model, in which a cosmological constant 
$\Lambda$ is added to the gravitational theory, predicting a constant energy density.
%, originated from the vacuum energy in quantum field theory.
Although this simplest model fits current cosmological observations very well, it faces several difficulties, such as the ``fine-tuning"~\cite{Review1, WBook} and ``coincidence''~\cite{Ostriker:1995rn, ArkaniHamed:2000tc} problems.

The running vacuum model (RVM) is one of the popular attempts to solve the latter problem~\cite{Ozer:1985ws, Carvalho:1991ut, Lima:1994gi, Lima:1995ea, Overduin:1998zv, Carneiro:2004iz, Shapiro:2009dh, Bauer:2005rpa, Dymnikova:2001ga, Alcaniz:2005dg, Barrow:2006hia, Lima:2015mca}.
In the RVM,  instead of a constant, $\Lambda$ is defined to be a function of the Hubble parameter $H$, and decays to matter (non-relativistic fluid) and radiation (relativistic fluid) in the evolution of the universe, leading to the same order of magnitude for the energy densities of dark energy and dark matter~\cite{Borges:2005qs, Borges:2007bh, Carneiro:2007bf, Zimdahl:2011ae, Alcaniz:2012mh, Geng:2016dqe, Sola:2015rra, Schutzhold:2002pr, Banerjee:2003fg, Klinkhamer:2009nn, Ohta:2010in, Cai:2010uf, Feng:2017usu, Li:2015vla, Li:2014eha, Li:2014cee}.
 Unlike the scalar tensor dark energy theory, 
 %{\color{red}
 such as the  simplest realistic scalar field dynamical one~\cite{Peebles:1987ek,Ratra:1987rm},
 %} 
 the RVM has no Lagrangian formula, indicating that this model is an effective theory from some
 other fundamental gravity theories.
 One possible origin of the RVM is from quantum effects induced by the cosmological renormalization group, 
resulting in $\Lambda = 3\nu H^2 + \Lambda_0$~\cite{Costa:2012xw, Sola:2014tta, Gomez-Valent:2014rxa, Shapiro:2004ch, EspanaBonet:2003vk, Tamayo:2015qla, Fritzsch:2016ewd, Basilakos:2009wi, Arcuri:1993pb,Sola:2013gha} with $\nu$ and $\Lambda_0$ constants.
It has been shown in Ref.~\cite{Sola:2016vis} that the RVM with $\nu=0$, i.e., the $\Lambda$CDM limit, is not favored by the observational data, whereas the best-fit for the model occurs at $\nu=4.8 \times 10^{-3}$, implying that the RVM with $\nu>0$ could well describe the evolution of our universe.
Results with similar constraints on $\nu$ have also been given in other types of the RVMs~\cite{Sola:2015wwa, Sola:2016jky, Sola:2016ecz, Gomez-Valent:2015pia, Gomez-Valent:2014fda}.
Clearly, it is interesting to investigate the matter power spectrum and cosmic microwave background radiation (CMB) temperature fluctuation in this scenario to see if the model is indeed better than the $\Lambda$CDM one.

In this paper, we will derive the growth equations of matter and radiation density fluctuations with the linear perturbation theory and illustrate the matter and CMB temperature power spectra,
which can significantly deviate from the $\Lambda$CDM prediction. 
We will show that the parameter $\nu$ will be further constrained when the observational data from ${\it Planck~2015}$ is taken into account.
Furthermore, one can also find at the end of this paper that the constraint on the RVM is about the same order of magnitude as
 that given in Ref.~\cite{Geng:2017apd}\footnote{In Ref.~\cite{Geng:2017apd}, a large number of data from $f(z) \sigma_8(z)$ and $H(z)$ observations as well as the CMB photon power spectrum have been included in the calculation.}, indicating that the CMB photon power spectrum provides a strong constraint on the RVM.

In principle,  dark energy has to be dynamical and its density fluctuation should be taken into account when the dark energy decay model is considered.
In order to investigate the dynamics of  dark energy, the running vacuum energy should be rewritten as a Lorentz scalar at the field equation level.
For example, in Refs.~\cite{Fabris:2006gt, Borges:2008ii, Geng:2016fql, VelasquezToribio:2009qp}, the cosmological constant is rewritten as $\Lambda = \Lambda(H)$ with $H=\nabla_{\mu} U^{\mu}/3$.
%However, the density fluctuation strongly interacts with the matter fluid at the scale deep inside the horizon, which conflicts with
 %the astrophysical observations.
In this work, we follow the perturbation method in Refs.~\cite{Borges:2008ii,  Basilakos:2014tha, Sola:2017jbl, Gomez-Valent:2018nib}, with which  dark energy simultaneously decays to  relativistic and non-relativistic matter and dilutes the density fluctuation.
%Besides, the expression of $H$ is not unique, and the physical behavior fully depends on its explicit expression.
%Due to these two reasons and without losing the generality, we consider  dark energy to be homogeneous and isotropic instead of a specific dark energy perturbation, i.e., $\delta \rho_{\Lambda} = 0$, where $\rho_{\Lambda}$ is the dark energy density.
%Note that the dark energy perturbation has been discussed in Refs.~\cite{Fabris:2006gt, Borges:2008ii, Geng:2016fql, VelasquezToribio:2009qp} by rewriting the Hubble parameter to be the Lorentz scalar of $\nabla_{\mu} U^{\mu}/3$ with $U^{\mu}=dx^{\mu}/ds$.
%In our study, we take dark energy to be homogeneous and isotropic instead of a specific dark energy perturbation.
%As a result, the particles, created from the dark energy decays, are homogeneously distributing to the universe.
We will also use the Markov chain Monte Carlo (MCMC) method to perform the global fit with the current observational data to further constrain the model.

This paper is organized as follows:
In Sec.~\ref{sec:model}, we introduce the RVM and review the background evolutions of matter, radiation and dark energy.
In Sec.~\ref{sec:perturbation}, we calculate the linear perturbation theory  and show the power spectra of the matter density
distribution and CMB temperature by the {\bf CAMB} program~\cite{Lewis:1999bs}.
In Sec.~\ref{sec:constraints}, we use the {\bf CosmoMC} package~\cite{Lewis:2002ah} to fit the model from the observational data.
% and give the constraints on cosmological parameters.
Our conclusions are presented  in Sec.~\ref{sec:conclusion}.

\section{Running vacuum model}
\label{sec:model}

The Einstein equation of the running vacuum model (RVM) is given by,
\begin{eqnarray}
\label{eq:fieldeq}
R_{\mu \nu} - \frac{g_{\mu \nu}}{2}R + \Lambda g_{\mu \nu} = \kappa^2 T_{\mu \nu}^M \,,
\end{eqnarray}
where $\kappa^2 = 8 \pi G$, $R=g^{\mu \nu} R_{\mu \nu}$ is the Ricci scalar, $\Lambda=\Lambda(H)$ is the time-dependent cosmological constant, and $T_{\mu \nu}^M$ is the energy-momentum tensor of matter and radiation.
In the Friedmann-Lema\"itre-Robertson-Walker (FLRW) metric of
%\begin{eqnarray}
%\label{eq:pert_metric0}
$ds^2 = a^2(\tau) [ -d\tau^2 + \delta_{ij} dx^i dx^j ]$, 
%\end{eqnarray}
the Friedmann equations are derived to be,
\begin{eqnarray}
\label{eq:Friedmann-1}
&& H^2= \frac{a^2\kappa^2}{3} \left( \rho_{M} + \rho_{\Lambda} \right) \,, \\
\label{eq:Friedmann-2}
&& \dot{H} = - \frac{a^2\kappa^2}{6} \left(  \rho_{M}+  3P_{M} + \rho_{\Lambda} + 3P_{\Lambda} \right) \,,
\end{eqnarray}
where $\tau$ is the conformal time, $H=da/(a d \tau)$ represents the Hubble parameter, $\rho_M=\rho_m+\rho_r$ ($P_{M}=P_m+P_r=P_r$) corresponds to the energy density (pressure) of matter and radiation, and $\rho_{\Lambda}$ ($P_{\Lambda}$) is the energy density (pressure) of the cosmological constant.
%, which can be defined f
 From Eq.~(\ref{eq:fieldeq}), we have
\begin{eqnarray}
\label{eq:rhol}
\rho_{\Lambda} = -P_{\Lambda} = \kappa^{-2} \Lambda(t) \,,
\end{eqnarray}
which leads to the equation of state (EoS) of $\Lambda$, given by
\begin{eqnarray}
\label{eq:wde}
w_{\Lambda} \equiv \frac{P_{\Lambda}}{\rho_{\Lambda}}= -1 \,.
\end{eqnarray}
In Eq.~(\ref{eq:fieldeq}), we consider $\Lambda$ to be a function of the Hubble parameter, 
given by~\cite{Costa:2012xw, Sola:2014tta, Gomez-Valent:2014rxa, Shapiro:2004ch, EspanaBonet:2003vk, Tamayo:2015qla, Fritzsch:2016ewd, Basilakos:2009wi, Arcuri:1993pb,Sola:2013gha}
\begin{eqnarray}
\label{eq:rnlam}
\Lambda = 3 \nu H^2 + \Lambda_0 \,,
\end{eqnarray}
where $\nu$ and $\Lambda_0$ are two free parameters.
%This model could originate from the quantum effect induced by the cosmological renormalization group running of the vacuum energy 
%in curved space-time~\cite{Sola:2013gha}.
In order to avoid the negative dark energy density in the early universe, we will concentrate on the RVM with $\nu \geq 0$ in our investigation.

Substituting Eq.~(\ref{eq:rnlam}) into the conservation equation, $\nabla^{\mu} (T^M_{\mu \nu}+T^\Lambda_{\mu \nu}) = 0$, we have
\begin{eqnarray}
\label{eq:contl}
\dot{\rho}_{\Lambda} + 3 H (1+w_{\Lambda}) \rho_{\Lambda} = 6 \nu H \dot{H} \neq 0 \,,
\end{eqnarray}
implying that dark energy unavoidably couples to matter  and radiation, given by
\begin{eqnarray}
\label{eq:contmr}
\dot{\rho}_l + 3 H (1+w_l) \rho_{a} = Q_l \,,
\end{eqnarray}
where $l$ represents matter ($m$) or radiation ($r$), $Q_l$ is the decay rate of the cosmological constant to $l=m$ or $r$, 
taken to be
\begin{eqnarray}
\label{eq:Qmr}
Q_l= -\frac{\dot{\rho}_{\Lambda}(\rho_{a}+P_l)}{\rho_M} = 3 \nu H (1+w_l) \rho_l \,,
\end{eqnarray}
and $w_{m(r)}=0$ ($1/3$) is the EoS of matter (radiation).
Subsequently, we  derive
\begin{eqnarray}
\label{eq:rhomr}
\rho_{a} = \rho_{a}^{(0)} a^{-3(1+w_{a})\xi} \,,
\end{eqnarray}
where $\xi=1-\nu$ and $\rho_{a}^{(0)}$ is the energy density of $a$ (matter or radiation) at $z=0$.

\section{Linear perturbation theory}
\label{sec:perturbation}

Since the RVM with the strong coupling $Q_l$, corresponding to $\nu \sim \mathcal{O}(1)$, is unable to  describe the evolution of the universe~\cite{Gomez-Valent:2014fda, Gomez-Valent:2015pia}, we only focus on the case of $\nu \ll 1$.
Note that $\nu$ has taken to be non-negative, i.e., $\nu \geq 0$, in order to avoid $\rho_{\Lambda}<0$ in the early universe.
The calculation follows the standard linear perturbation theory with the synchronous gauge~\cite{Ma:1995ey}.
The metric is given by,
\begin{eqnarray}
\label{eq:pert_metric}
ds^2 = a^2(\tau) \left[ -d\tau^2 + (\delta_{ij} + h_{ij}) dx^i dx^j \right] \,,
\end{eqnarray}
where
\begin{eqnarray}
\label{eq:pert_h}
h_{ij} = \int d^3 k e^{i \vec{k} \cdot \vec{x}} \left[ \hat{k}_i \hat{k}_j h(\vec{k},\tau) + 6 \left( \hat{k}_i \hat{k}_j -\frac{1}{3} \delta_{ij} \right) \eta(\vec{k},\tau) \right] \,,
\end{eqnarray}
$i,j=1,2,3$, $h$ and $\eta$ are two scalar perturbations in the synchronous gauge, and $\hat{k} = \vec{k}/ k $ is the k-space unit vector.
From $\nabla^{\mu} (T^M_{\mu \nu}+T^\Lambda_{\mu \nu}) = 0$ with $\delta
 T^0_0 = \delta \rho_M$, $\delta
 T^0_i = -T^i_0 = (\rho_M+P_M) v^i_M$ and $\delta
 T^i_j = \delta P_M \delta^i_j$, one can obtain the matter and radiation density perturbations, given by~\cite{Borges:2008ii, Basilakos:2014tha, Sola:2017jbl, Gomez-Valent:2018nib},
\begin{eqnarray}
\label{eq:drho}
&&\dot{\delta}_l = - (1+w_l)\left( \theta_l + \frac{\dot{h}}{2} \right) - 3 H \left( \frac{\delta P_l}{\delta \rho_l} - w_l \right) \delta_l -\frac{a Q_l}{\rho_l} \delta_l \,, \\
\label{eq:dtheta}
&&\dot{\theta}_l = -H\left( 1-3w_l \right) \theta_l - \frac{\dot{w_l}}{1+w_l} \theta_l + \frac{\delta P_l / \delta \rho_l}{1+w_l} \frac{k^2}{a^2} \delta_l - \frac{a Q_l}{\rho_l} \theta_l \,,
\end{eqnarray}
where $\delta_l \equiv \delta \rho_l / \rho_l$ and $\theta_l = ik_i v^i_l$  are the density fluctuation and the divergence of fluid velocity, respectively.
Note that Eqs.~\eqref{eq:drho} and \eqref{eq:dtheta} describe the evolutions of density fluctuations of the perfect fluids without  interactions between them.
If the interactions between any two fluids are taken into account, these equations should be further modified.
Taking the photon-proton interaction to be the example, one has to add the additional term, $a n_e \sigma_T (\theta_b - \theta_{\gamma})$, at the RHS of Eq.~\eqref{eq:dtheta} when $l=\gamma$, and the details of the equations can be found in Ref.~\cite{Ma:1995ey}.
%Secondly, as mentioned in the introduction that we have taken $\delta_{\Lambda} =0$, so the particle creation from the dark energy decay homogeneously distributes to the universe, smoothing the density fluctuation of the matter and radiation. 
In Eqs.~\eqref{eq:drho} and \eqref{eq:dtheta}, one can observe that the last terms in the two equation slow down the growths of $\delta_l$ and $\theta_l$ if $\nu$ in Eq.~\eqref{eq:Qmr} is positive.

To show how the running vacuum scenario in Eq.~(\ref{eq:rnlam}) influences the physical observables, 
we use the open-source program {\bf CAMB}~\cite{Lewis:1999bs}, 
in which we modify the background density evolutions and the evolution equations of $\delta$ and $\theta$ in terms of Eqs.~(\ref{eq:rhomr}), (\ref{eq:drho}) and (\ref{eq:dtheta}).
By taking $1 \gg \nu \geq 0$, most of particles are created at the end of inflation, whereas the energy density from the dark energy decay is tiny, hinting that the RVM shares the same initial condition as that in the $\Lambda$CDM model.
In addition, the matter-radiation equality $z_{eq}$ slightly changes, given by
\begin{eqnarray}
\left. \frac{\rho_m(z)}{\rho_r(z)} \right\vert_{z=z_{eq}} = 1 \,.
\end{eqnarray}

\begin{figure}
\centering
\includegraphics[width=0.49 \linewidth]{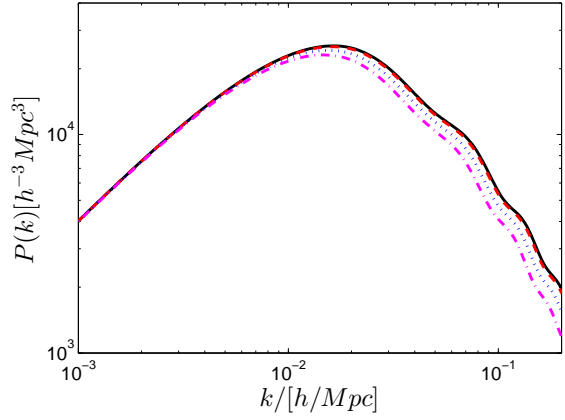}
\caption{
The matter power spectrum $P(k)$ as a function of the wavelength $k$ with $\nu=0$ (solid line), $10^{-3}$ (dashed line), $5 \times 10^{-3}$ (dotted line) and $10^{-2}$ (dash-dotted line), where the boundary conditions are taken to be $\Omega_b h^2=2.23\times 10^{-2}$, $\Omega_c h^2=0.118$, $h=0.68$,  $A_s=2.15 \times 10^{-9}$, $n_s=0.97$, $\tau=0.07$ and $ \Sigma m_{\nu}=0.06$~eV, respectively.
}
\label{fg:1}
\end{figure}

In Fig.~\ref{fg:1}, we present the matter power spectrum $P(k) \sim \langle \delta_m^2(k) \rangle$ as a function of the wavenumber $k$ with $\nu=0$ (solid line), $10^{-3}$ (dashed line), $5 \times 10^{-3}$ (dotted line) and $10^{-2}$ (dash-dotted line).
As discussed earlier in this section, the matter density fluctuation is diluted by the creation of particles, so that 
the results of $P(k)$ at large $k$ and $\nu$ in the RVM  significantly deviate from that in the $\Lambda$CDM model (solid line).
%The larger value of $\nu$ and $k$ are, the more significant suppressions for $P(k)$ are.

\begin{figure}
\centering
\includegraphics[width=0.49 \linewidth]{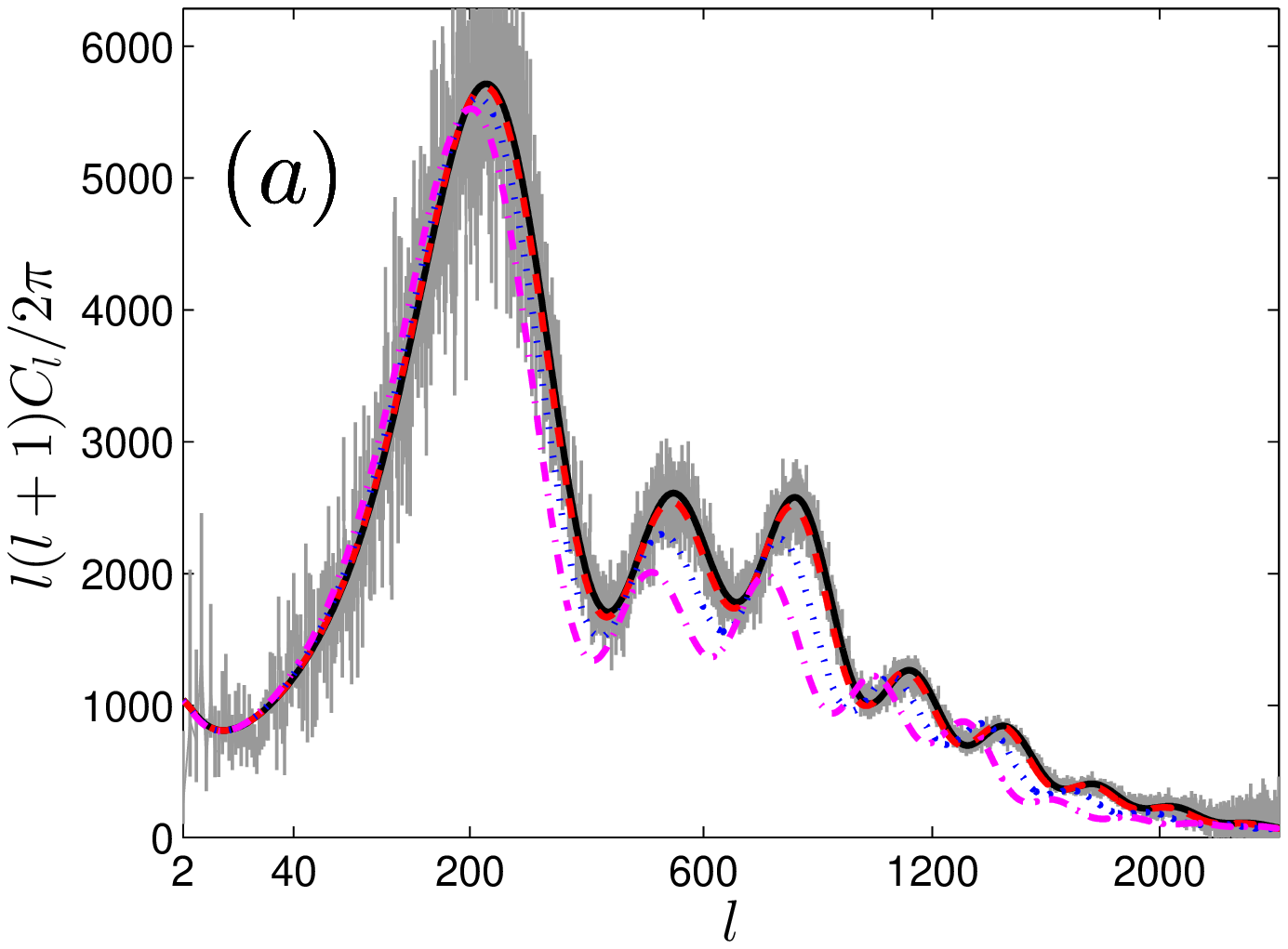}
\includegraphics[width=0.49 \linewidth]{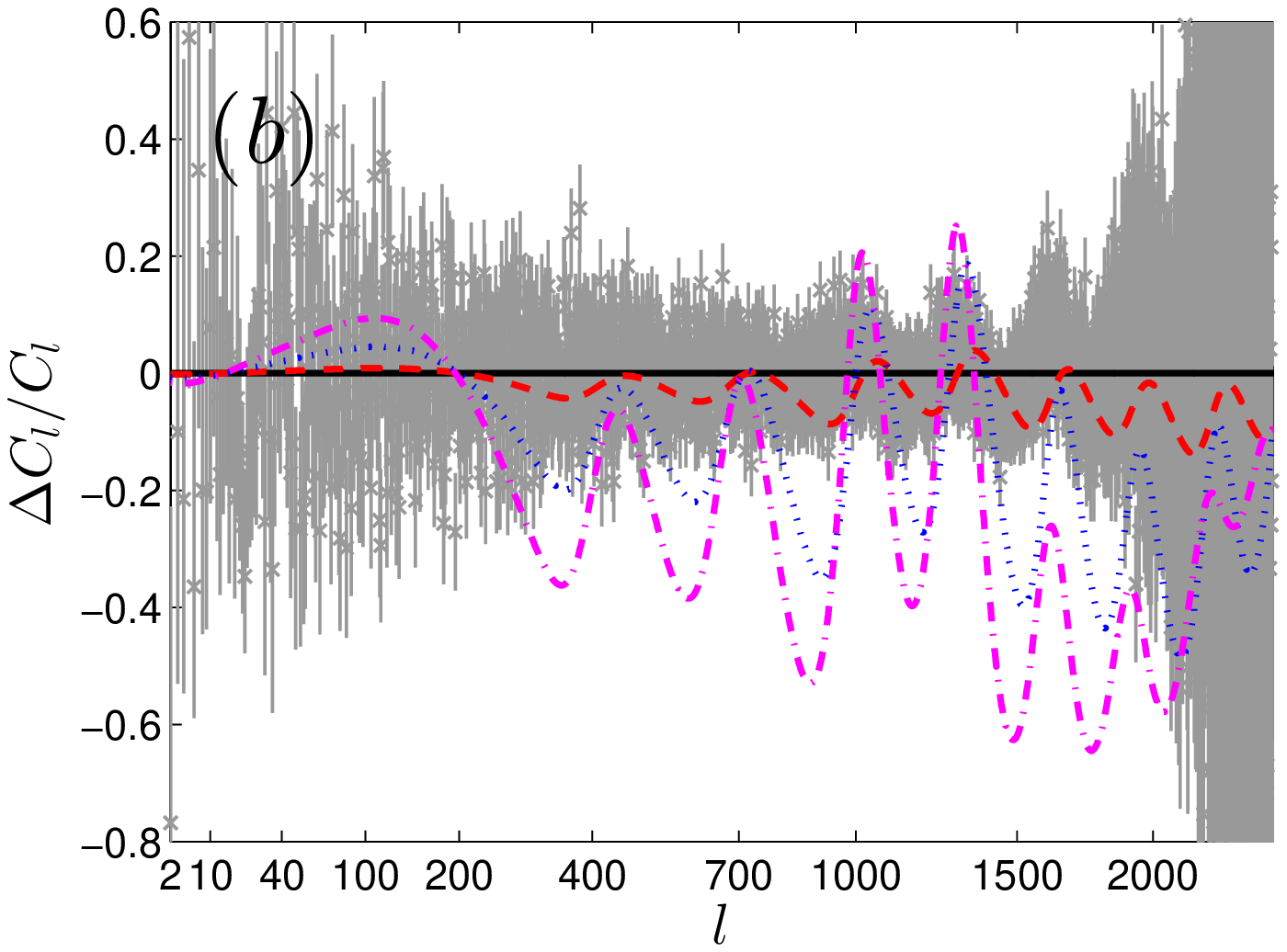}
\caption{
The CMB temperature power spectra of (a) $l(l+1)C_l/2\pi$  and (b) $\Delta C_l/C_l=(C_l^{RVM}-C_l^{\Lambda CDM})/C_l^{\Lambda CDM}$ with $T=2.73~K$, where legend is the same as Fig.~\ref{fg:1}  and the grey points are the unbinned TT mode data from the {\it Planck 2015}.
}
\label{fg:2}
\end{figure}
Fig.~\ref{fg:2} shows the CMB temperature spectra of (a)  $l(l+1)C_l/2\pi$  and (b) $\Delta C_l/C_l=(C_l^{RVM}-C_l^{\Lambda CDM})/C_l^{\Lambda CDM}$ in the RVM with $\nu=0$ (solid line), $10^{-3}$ (dashed line), $5 \times 10^{-3}$ (dotted line) and $10^{-2}$ (dash-dotted line), 
 where the grey points are the unbinned TT mode data from the {\it Planck 2015}.
We see that  the CMB temperature spectra are significantly suppressed in the RVM. 
%The deviation of $C_l$ from that in the $\Lambda$CDM model can be larger than 10\% with $\nu \gtrsim O(10^{-3})$.
The maximum deviations of $C_l$ from that in the $\Lambda$CDM model can be 13.8\%, 48.6\% and 64.5\% with $\nu = 10^{-3}$, $5\times 10^{-2}$ and $10^{-2}$, respectively.
Due to the accurate measurement from the {\it Planck 2015}, we can estimate that the allowed  range of $\nu$ should be at the same order of or less than $O(10^{-3})$.
%{\color{red}
We note that
it is important to note there is a degeneracy with spatial curvature 
when studying dynamical dark energy models.
Clearly, it might not be reasonable to retain flat geometry if one wants to get
a realistic set of observational constraints.
As shown in the literature~\cite{Hu:2001bc, Ooba:2017npx}, a positive spatial curvature shifts the CMB temperature spectra to the smaller $l$ and rises $C_l$
in the small $l$ region.
The former phenomenon degenerate with our RVM, whereas the later case is not, i.e., the RVM moves the high-$l$ and 
keeps the low-$l$ spectra in Fig.~\ref{fg:2}a.
Moreover, as pointed out in Ref.~\cite{Farooq:2013dra}, when
curvature is allowed to be a free parameter, the constraints on dark energy
dynamics weaken considerably.
In this work, we are interested in  the curvature-free case and leave the discussion of the spatial curvature to the future work.
%}

\section{Observational constraints}
\label{sec:constraints}

\begin{table}[ht]
\begin{center}
\caption{ Priors for cosmological parameters with $\Lambda= 3\nu H^2 + \Lambda_0$.  }
\begin{tabular}{|c||c|} \hline
Parameter & Prior
\\ \hline
Model parameter $\nu$& $0 \leq \nu \leq 3 \times 10^{-4}$
\\ \hline
Baryon density & $0.5 \leq 100\Omega_bh^2 \leq 10$
\\ \hline
CDM density & $10^{-3} \leq \Omega_ch^2 \leq 0.99$
\\ \hline
Optical depth & $0.01 \leq \tau \leq 0.8$
\\ \hline
Neutrino mass sum& $0 \leq \Sigma m_{\nu} \leq 2$~eV
\\ \hline
 $\frac{\mathrm{Sound \ horizon}}{\mathrm{Angular \ diameter \ distance}}$  & $0.5 \leq 100 \theta_{MC} \leq 10$
\\ \hline
 Scalar power spectrum amplitude & $2 \leq \ln \left( 10^{10} A_s \right) \leq 4$
\\ \hline
 Spectral index & $0.8 \leq n_s \leq 1.2$
\\ \hline
\end{tabular}
%\vskip 0.2in
\label{tab:1}
\end{center}
\end{table}

\begin{figure}
\centering
\includegraphics[width=0.98 \linewidth]{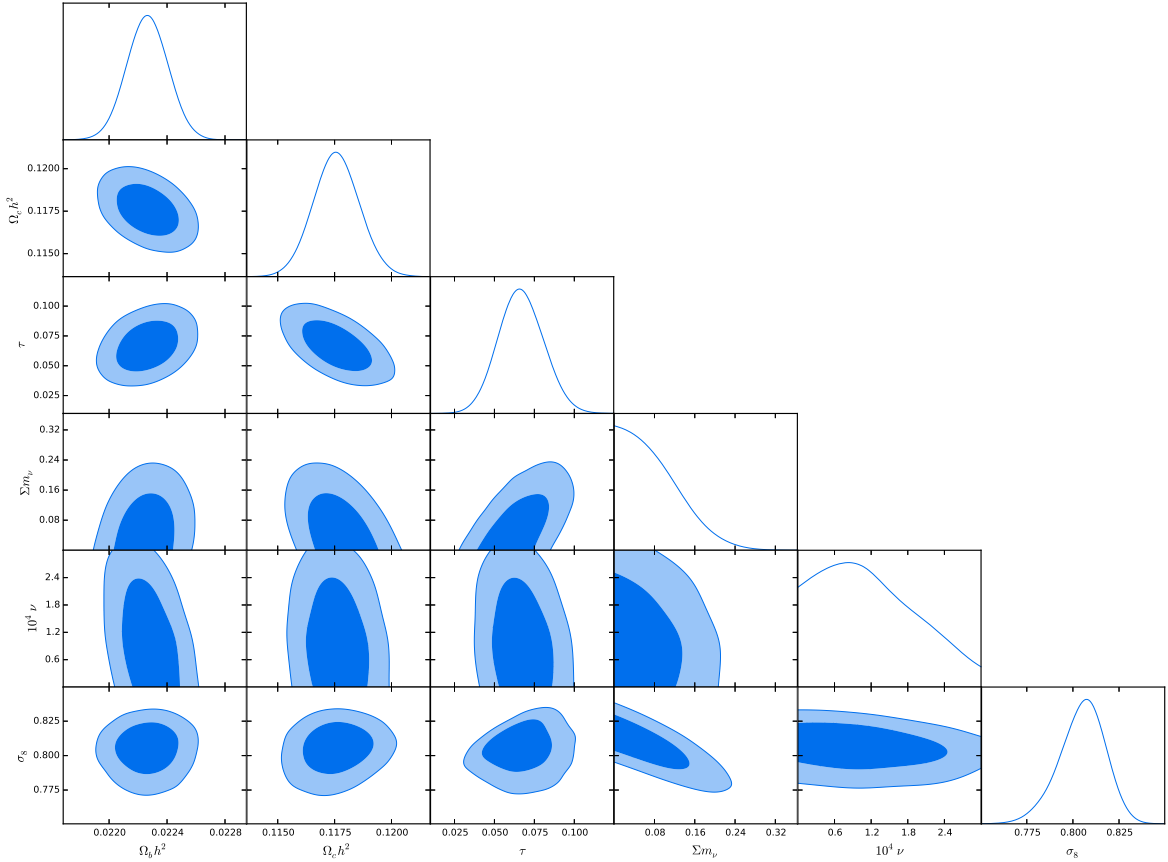}
\caption{
One and two-dimensional distributions of  $\Omega_bh^2$, $\Omega_ch^2$, $\tau$,  $\Sigma m_{\nu}$, $\nu$ and $\sigma_8$, where the contour lines represent  68\% and 95\% confidence levels, respectively. }
\label{fg:3}
\end{figure}

We now perform the open-source {\bf CosmoMC} program~\cite{Lewis:2002ah} with the MCMC method to explore 
a more precise range for the model parameter $\nu$.
% as well as the cosmological observables.
The dataset includes the cosmic microwave background radiation (CMBR), combined with the CMB lensing, from {\it Planck 2015} TT, TE, EE, low-$l$ polarization~\cite{Adam:2015wua,Aghanim:2015xee,Ade:2015zua}; baryon acoustic oscillation (BAO) data from 6dF Galaxy Survey~\cite{Beutler:2011hx}, SDSS DR7~\cite{Ross:2014qpa} and BOSS~\cite{Gil-Marin:2015nqa}; matter power spectrum data from SDSS DR4 and WiggleZ~\cite{Blake:2010xz, Blake:2011en, Parkinson:2012vd}, and weak lensing data from CFHTLenS~\cite{Heymans:2013fya}.
The priors of the various parameters are listed in Table.~\ref{tab:1}.

\begin{table}[ht]
\begin{center}
\caption{Fitting results for the RVM with $\Lambda = 3\nu H^2 + \Lambda_0$ and $\Lambda$CDM, where $\chi^2_{\mathrm{Best-fit}}=\chi^2_{\mathrm{CMB}}+\chi^2_{\mathrm{BAO}}+\chi^2_{\mathrm{MPK}}+\chi^2_{\mathrm{lensing}}$, and limits are given at $95\%$ confidence level ($\nu$ is calculated within $68\%$ C.L.).} 
\begin{tabular}{|c||c|c|} \hline
Parameter &  RVM & $\Lambda$CDM
\\ \hline
Model parameter ($10^4 \nu$) & $1.19^{+0.35}_{-1.19}$ ($68\%$ C.L.)& --
\\ \hline
Baryon density ($ 100 \Omega_bh^2$) & $ 2.23^{+0.02}_{-0.03}$  & $ 2.23 \pm 0.03$
\\ \hline
CDM density ($ \Omega_ch^2 $) & $ 0.118 \pm 0.002$ & $ 0.118 \pm 0.002$
\\ \hline
Matter density ($ \Omega_m $) & $ 0.308^{+0.015}_{-0.013}$ & $ 0.306 \pm 0.014$
\\ \hline
 Hubble parameter ($ H_0$) $(km/s \cdot Mpc)$ & $67.58^{+1.14}_{-1.23}$ & $ 67.87^{+1.07}_{-1.22}$
\\ \hline
Optical depth ($ \tau$) & $  6.66^{+2.82}_{-2.68} \times 10^{-2} $ & $ 6.99^{+2.83}_{-2.77}\times 10^{-2} $
\\ \hline
Neutrino mass sum ($\Sigma m_{\nu} $) & $< 0.186$~eV & $< 0.200$~eV
\\ \hline
 $100 \theta_{MC}$ & $ 1.0411 \pm 0.0006$ & $ 1.0409 \pm 0.0006$
\\ \hline
 $\ln \left( 10^{10} A_s \right)$ & $ 3.06^{+0.06}_{-0.05}$ & $ 3.07 \pm 0.05$
\\ \hline
 $n_s$ & $ 0.970^{+0.007}_{-0.008}$ & $ 0.970^{+0.007}_{-0.008}$
\\ \hline
$\sigma_8$ & $ 0.805^{+0.023}_{-0.027}$ & $ 0.808^{+0.025}_{-0.026}$
\\ \hline
$z_{eq}$  & $3345^{+46}_{-44}$ & $3348 ^{+45}_{-46} $
\\ \hline
 $\chi^2_{\mathrm{Best-fit}}$ & $ 13968.8$ & $ 13969.8$
\\ \hline
\end{tabular}
\label{tab:2}
\end{center}
\end{table}

In Fig.~\ref{fg:3}, we show the global fit from the observational data.
In Table.~\ref{tab:2}, we list the allowed ranges for various cosmological parameters at $95\%$ confidence level ($\nu$ at $68\%$ one).
We find that the best-fit occurs at $\nu = 1.19 \times 10^{-4}$ with $\chi^2_{\mathrm{RVM}}=13968.8$, which is smaller than 
$\chi^2_{\mathrm{\Lambda CDM}}=13969.8$ in the $\Lambda$CDM model. 
This result demonstrates that the RVM with $\Lambda= 3 \nu H^2 + \Lambda_0$ is preferred by the cosmological observations, in which
$\nu \lesssim 1.54 \times 10^{-4}$ is constrained at $68\%$ confidence level.
However,  the model is unable to be distinguished from the $\Lambda$CDM model within $1 \sigma$ confidence level.
%{\color{red}
In addition,
although our result of $\chi^2$ is smaller than that in $\Lambda$CDM, it is clearly not significant due
to the large overall values of $\chi^2$ for both models.
%}
Comparing to the best fitted value of $\nu = 4.8\times 10^{-3}$ in Ref.~\cite{Sola:2016vis}, our simulation further lowers 
the model parameter $\nu$ more than one order of magnitude.

\section{Conclusions}
\label{sec:conclusion}

We have studied the RVM with $\Lambda= 3 \nu H^2 + \Lambda_0$, in which dark energy decays to both matter and radiation.
We have calculated the evolution equations of the matter density fluctuation $\delta(k,a)$ and the divergence of the fluid velocity $\theta(k,a)$ with the linear perturbation theory.
We have shown that the decaying dark energy suppresses both $\delta$ and $\theta$, while the power spectra of the matter density distribution and CMB temperature fluctuation significantly deviate from those in the $\Lambda$CDM model.
By performing the global fit from the cosmological observations, we have obtained that $\nu \leq 1.54 \times 10^{-4}$ with the best-fit $\chi^2_{\mathrm{RVM}} - \chi^2_{\Lambda \mathrm{CDM}} = -1.0$ at $\nu=1.19 \times 10^{-4}$.
Such a strong constraint on $\nu$ with a small $\chi^2$ difference  is due to the TT mode of the CMB measurement, which only allows the physical observables of the modified gravity models slightly deviating from that of the $\Lambda$CDM model.
This situation can be seen in not only the RVM but also some other dark energy models, such as XCDM and $\phi$CDM~\cite{Ooba:2018dzf, Park:2018bwy, Park:2018fxx}.
In summary, although the RVM perfectly describes the evolution of our universe, the accurate measurement from the {\it Planck 2015} strongly constrains this scenario,
and the allowed window of the model parameter $\nu$ is an order of magnitude smaller than those results in Refs.~\cite{Sola:2015wwa, Sola:2016jky, Sola:2016ecz, Gomez-Valent:2015pia, Gomez-Valent:2014fda}.
It is clear that the $\Lambda$CDM model is hardly to be ruled out by the current cosmological observations yet.
%{\color{red}
Finally, it is interesting to mention that the results from 
 the spatially-flat XCDM and $\phi$CDM models~\cite{Ooba:2018dzf, Park:2018bwy, Park:2018fxx} 
also do not agree with those in Refs.~\cite{Sola:2015wwa, Sola:2016jky, Sola:2016ecz, Gomez-Valent:2015pia, Gomez-Valent:2014fda}.
%}

\section*{Acknowledgments}
We would like to thank Professor Joan Sola 
%(Barcelona U. \& ICC, Barcelona U.) 
for helpful discussions.
This work was partially supported by National Center for Theoretical Sciences, MoST (MoST-104-2112-M-007-003-MY3, MoST-107-2119-M-007-013-MY3
 and MoST-106-2917-I-564-055) and the Newton International Fellowship (NF160058) from the Royal Society (UK).

\end{document}